\documentclass{article}
\usepackage{spconf,amsmath,graphicx}

\usepackage{hyperref}
\usepackage{xcolor}
\usepackage{url}
\usepackage{subcaption}
\usepackage{booktabs} 

\newcommand*\samethanks[1][\value{footnote}]{\footnotemark[#1]}

\title{TOWARDS IDENTITY PRESERVING NORMAL TO DYSARTHRIC VOICE CONVERSION} 
\name{Wen-Chin Huang$^1$*\thanks{*Equal contribution.}, Bence Mark Halpern$^{2,3,4}$*\samethanks[1], Lester Phillip Violeta$^1$, Odette Scharenborg$^2$, Tomoki Toda$^1$}
\address{$^1$Nagoya University, Japan\\
         $^2$Multimedia Computing Group, Delft University of Technology, Delft, The Netherlands\\
         $^3$University of Amsterdam, Amsterdam, The Netherlands\\
         $^4$Netherlands Cancer Institute, Amsterdam, The Netherlands}

\begin{document}
\ninept
\maketitle
\begin{abstract}
We present a voice conversion framework that converts normal speech into dysarthric speech while preserving the speaker identity. Such a framework is essential for (1) clinical decision making processes and alleviation of patient stress, (2) data augmentation for dysarthric speech recognition. This is an especially challenging task since the converted samples should capture the severity of dysarthric speech while being highly natural and possessing the speaker identity of the normal speaker. To this end, we adopted a two-stage framework, which consists of a sequence-to-sequence model and a nonparallel frame-wise model. Objective and subjective evaluations were conducted on the UASpeech dataset, and results showed that the method was able to yield reasonable naturalness and capture severity aspects of the pathological speech. On the other hand, the similarity to the normal source speaker's voice was limited and requires further improvements.
\end{abstract}
\begin{keywords}
voice conversion, pathological speech, dysarthric speech, sequence-to-sequence modeling, autoencoder
\end{keywords}
\section{Introduction}
\label{sec:intro}

Neural voice conversion  (VC) has substantially improved the naturalness of synthesized speech in a wide range of tasks, including read speech \cite{VTN}, emotional speech \cite{zhou20_odyssey} and whispered speech \cite{cotescu2019voice}. However, pathological VC (and TTS too) is a largely unexplored area, which has several interesting applications. In this work, we focus on normal-to-dysarthric (N2D) VC, which refers to the task of converting normal speech to dysarthric speech. N2D VC could be applied in informed decision making related to the medical conditions at the root of the speech pathology. For instance, an oral cancer surgery results in changes to a speaker's voice. The availability of a VC model that can generate how the voice could sound after surgery could help the patients and clinicians make informed decisions about the surgery and alleviate the stress of the patients. Another application is the improvement of automatic speech recognition (ASR) by augmenting the training dataset with additional pathological data. Such augmentation could ease the low-resource constraints of a pathological ASR task. 

In addition to the requirements for conventional VC, N2D VC has its own unique requirements, each corresponding to one research question:

\noindent\textbf{RQ1: Do the converted samples sound as natural as real dysarthric samples?} Naturalness is a basic requirement in all speech synthesis tasks, but it becomes challenging under the context of N2D VC  because listeners seem to confuse  \textit{naturalness} and \textit{severity} \cite{illa21_ssw}.

\noindent\textbf{RQ2: Is the VC model able to retain the speaker identity of the source normal speaker?} Since it is often impossible to collect ground truth pathological speech data of a normal source speaker, training a VC model that directly maps a normal source speech to its pathological counterpart is infeasible. Thus, specific techniques need to be developed to tackle this issue.
In addition, evaluation of similarity is hard because listeners have to determine the similarity of a converted pathological speech to the source speaker while having access to only a normal speech of him/her.


\noindent\textbf{RQ3: Is the VC model able to model severity characteristics in a linear way, so that expert listeners perceive more severe samples as more severe?} As the condition of patients deteriorates, the severity of the patient's voice will increase. To capture the progress, it is essential to correctly model the severity of the converted speech. This requires modifying specific attributes of speech, such as speaking rate and insertion of pauses.



\begin{figure}[t]
	\centering
	
	\begin{subfigure}[b]{\columnwidth}
		\centering
  		\includegraphics[width=\textwidth]{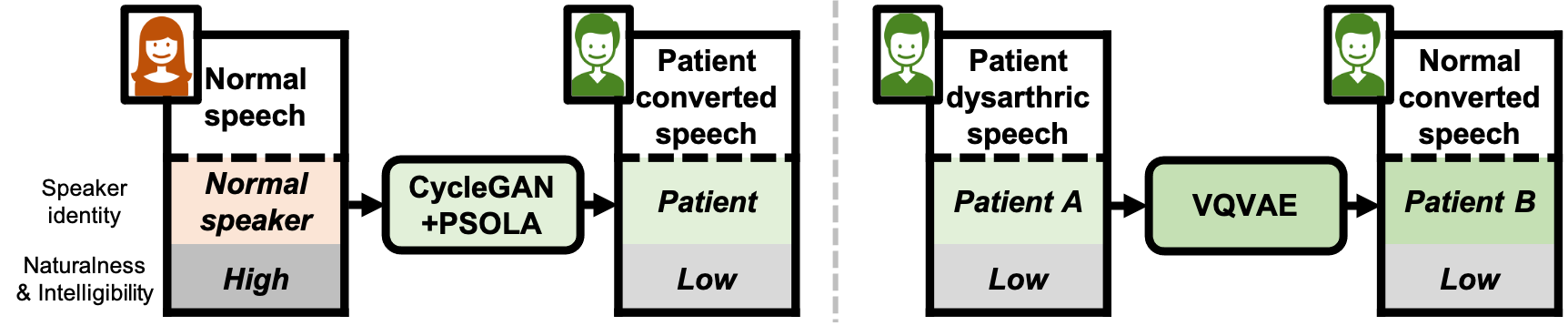}
		\caption{Previous works. Left: \cite{halpern2021objective}. Right: \cite{illa21_ssw}}
   		\label{fig:previous-works}
	\end{subfigure}\\
	
	\vspace{0.3cm}
	
	\begin{subfigure}[b]{\columnwidth}
		\centering
	    \includegraphics[width=\textwidth]{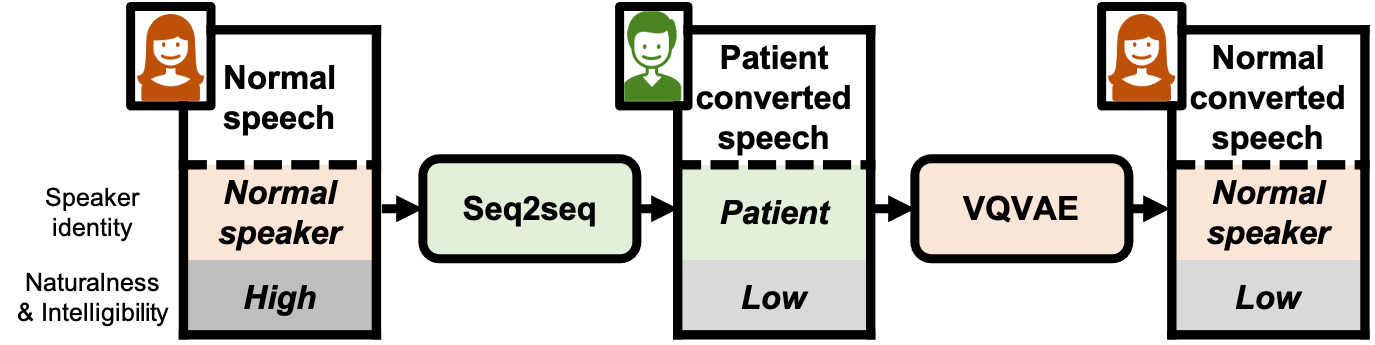}
		\caption{Proposed two-stage approach.}
   		\label{fig:proposed}
   	\end{subfigure}\\
	
	\centering
	\caption{Illustration of previous work and the proposed method for N2D VC.}
	\vspace{-0.5cm}
	\label{fig:methods}
\end{figure}

In this work, we aim to create an identity preserving N2D VC system. The key advantage of this approach is that it allows arbitrary inputs from the source normal speaker, while preserving its identity. The aim of this work is to evaluate the model in a more practical setting than \cite{illa21_ssw} by taking normal speech as input, which alleviates the need of maintaining a pathological voice bank described there. 
Inspired by \cite{dvc-vtn-vae}, the proposed method is a two-stage approach, as depicted in Figure~\ref{fig:proposed}. In the first stage, to capture the unique temporal structure of dysarthric speech, we adopt the Voice Transformer Network (VTN) \cite{VTN, VTN-TASLP}, a sequence-to-sequence (seq2seq) VC model based on the Transformer \cite{transformer} architecture. The converted speech at this stage has the characteristics of dysarthric speech, with an unwanted speaker identity of the reference dysarthric speaker. Then, the normal source speaker identity is restored through a frame-by-frame autoencoder-based VC model \cite{crank}, which is assumed to be able to preserve local speech attributes related to dysarthria. We evaluated the proposed method on UASpeech \cite{UASpeech}, and the method achieves good naturalness results, is able to mimic the severity of pathological speech according to three speech language pathologists, while having limited ability to preserve the source speaker's characteristics.


\section{Related Works}

\subsection{Normal-to-dysarthric VC for data augmentation in ASR}
\label{ssec:n2d-vc-for-asr}

Previous research on data augmentation for dysarthric speech has shown promising improvements in ASR word error rates.
The mainstream is to use frame-wise models such as deep convolutional general adversarial networks (DCGANs) \cite{asr-gan} or Transformer Encoders \cite{asr-vocab} to convert the speech timbre. As these models do not change the length, extra procedures are needed to change the speaking rate, including speed perturbation \cite{asr-gan} or dynamic time warping \cite{asr-vocab}.

There are several downsides to this line of work.
As ASR only requires the various dysarthric features to be modeled, the speaker identity of the normal speaker is not retained after conversion. Also, no evaluation methods were conducted to measure the severity of the samples, which means that it was not verified whether the proposed methods were truly able to model the dysarthric features well.
In this work, we use a seq2seq model to jointly convert the timbre and speaking rate, which was shown to be more effective than converting them separately in conventional VC \cite{S2S-iFLYTEK-VC}. We also address the identity preservation issue with the proposed two-stage approach and conduct subjective evaluations to verify if the severity is indeed modeled.



\subsection{Normal-to-dysarthric VC for clinical usage}

There are two previous works that focus on VC for clinical usage. 
The diagram on the left of Figure~\ref{fig:previous-works} depicts an N2D VC system presented in \cite{halpern2021objective}, which was a combination of a CycleGAN-based frame-wise VC model and a PSOLA-based speech rate modification process. This method suffers from the same issues as those in Section~\ref{ssec:n2d-vc-for-asr}, including audible vocoder artifacts brought by the extra PSOLA operation, and the inability to preserve the speaker identity of the control speaker.

A different work \cite{illa21_ssw} is depicted on the righthand side of Figure~\ref{fig:previous-works}. The authors focused on dysarthric-to-dysarthric VC, by using a frame-wise VC model called HL-VQ-VAE \cite{HL-VQ-VAE}. 
However, the setup was not flexible in that (1) a severity-matched VC setup was required to avoid the need of varying speech rates, and (2) the method required a pathological source utterance, wherein real-world applications we might want to synthesize an arbitrary utterance from the normal source speaker.



\section{Proposed Framework}

Given a speech sample from a normal speaker, N2D VC aims to change the characteristics into that of a dysarthric speech, while preserving the speaker identity of the source normal speaker. In the following subsections, we describe the two components, the parallel seq2seq model and the nonparallel frame-wise model, of our proposed two-stage approach for N2D VC in detail.

\subsection{Many-to-one seq2seq modeling}

The goal in the first stage is to completely capture the characteristics of the dysarthric speech. Following \cite{dvc-vtn-vae}, we adopted the VTN \cite{VTN, VTN-TASLP}, a Transformer-based \cite{transformer} seq2seq model tailored for VC. When a parallel corpus is available, seq2seq modeling is considered state-of-the-art due to its ability to convert the prosodic structures in speech, which is critical in N2D VC. 
However, collecting a parallel corpus is especially difficult in our case since it is impractical (almost not feasible) to collect a large amount of data from dysarthric patients. To solve the data scarcity problem, we applied two techniques, as described below.

First, a TTS pretraining technique is applied which facilitates the core ability of a seq2seq VC model, i.e., encode linguistic-rich hidden representations by pretraining using a large-scale TTS dataset \cite{VTN, VTN-TASLP}. This technique is flexible in that the VC corpus and the pretraining TTS dataset can be completely different in terms of speaker and content, even when trained between normal and dysarthric speakers. In \cite{dvc-vtn-vae}, it was shown that training using only 15 minutes of speech from each speaker can yield good results.

Second, we trained the VTN in a many-to-one (referred to as M2O) fashion. Considering that it is easier to collect data from normal speakers rather than patients, we assume that apart from the data of the source normal speaker, we also have access to a set of parallel training set from multiple normal speakers. Given a training utterance from any of the normal speakers, the VTN model is trained to convert to the predefined target dysarthric speaker.
M2O training was also used in \cite{parrotron}, except they used an auxiliary phoneme recognition regularization loss.

\subsection{Nonparallel frame-wise model}
\label{ssec:vae}

In the second stage, given the converted dysarthric speech, the goal is to restore the identity of the source normal speaker while preserving the dysarthric attributes. We adopted the same assumption as in \cite{dvc-vtn-vae}: a nonparallel frame-wise VC model changes only time-invariant characteristics such as the speaker identity, while preserving time-variant characteristics, such as the pronunciation. As in \cite{dvc-vtn-vae}, we used \textit{crank} \cite{crank}, an open-source VC software that combines recent advances in VQVAE \cite{VQVAE}-based VC methods, including the use of hierarchical architectures, cyclic loss and adversarial training, to carry out the conversion of the speaker identity step. For the remainder of this paper, we refer to this model as \textit{VAE} for short. 



\section{Experimental setup}

\subsection{Dataset}
\label{ssec:dataset}

We used the UASpeech dataset \cite{UASpeech}, which contains parallel word recordings of 15 dysarthric speakers and 13 normal control speakers. The training and test set consist of 510 and 255 utterances, respectively. Each dysarthric speaker is categorized to one of three intelligibility groups: low, mid, and high, which correspond to $0-25\%$, $25-75\%$, and $75-100\%$ subjective human transcription error rate (STER). The intelligibility of each speaker was judged by 5 non-expert American English native speakers. We chose two dysarthric speakers from each intelligibility group (high: M08, M10; mid: M05, M11; low: M04, M12) as test speakers for VC. For each dysarthric speaker, a separate VTN was trained using the data of that speaker and all control speakers. For the VAE model, in our preliminary experiments, we found that it was crucial to train with only the normal data rather than training with a mix of dysarthric and normal datasets. We thus used data from the 13 control speakers only.

\subsection{Implementation}

The implementation of the VTN (the left rounded rectangle in Figure~\ref{fig:proposed}) was based on the open-source toolkit ESPnet \cite{espnet, espnet-2020}. The detailed configuration can be found online\footnote{\url{https://github.com/espnet/espnet/tree/master/egs/arctic/vc1}}. The TTS pretraining was conducted with M-AILABS judy \cite{M-AILABS}, which was 31 hr long.
\textit{crank}, which we base our implementation of VAE on (the right rounded rectangle in Figure~\ref{fig:proposed}), is also open-sourced and can be accessed freely\footnote{\url{https://github.com/k2kobayashi/crank}}. 
Parallel WaveGAN (PWG) \cite{parallel-wavegan} was used as the neural vocoder. We followed an open-source implementation\footnote{\url{https://github.com/kan-bayashi/ParallelWaveGAN}}. The training data of PWG contained the audio recordings of all control speakers in UASpeech.

\vspace{-0.2cm}
\subsection{Objective evaluation metrics}

The speech sample outputs of the two stages (VTN, VTN-VAE) are separately evaluated using the metrics described in this section, whenever the evaluation does not require ground truth. In this evaluation, we considered conversion pairs between all 13 normal source speakers and the 6 dysarthric speakers mentioned in Section~\ref{ssec:dataset}.




\subsubsection{P-ESTOI/P-STOI}


P-ESTOI/P-STOI were previously demonstrated to work well for the objective evaluation of dysarthric speech \cite{janbakhshi2019pathological}. These methods focus on quantifying distortion in the time-frequency structure of the speech signal, which is related to severity and naturalness (RQ1 and RQ3).
In short, we used multiple gender-specific ground truth control utterances to form a reference utterance.
By calculating the frame-level cross-correlation of each pathological utterance with the reference utterance, we obtain an utterance-level P-ESTOI/P-STOI score. Taking the mean of each utterance-level score, we obtain a speaker-level score,  which is correlated with the STER scores for the six speakers to obtain $r$. This is repeated with the ground truth speakers to obtain $r_{GT}$

\subsubsection{Phoneme error rate}

The phoneme error rate (PER) calculated with a phoneme recognizer evaluates the intelligibility, which is also related to severity and naturalness (RQ1 and RQ3). We use a pre-trained Kaldi ASR model with the same specifications as the one used in \cite{purohit2020intelligibility} for phoneme recognition. The ASR was trained with the TIMIT dataset and used an HMM acoustic model. The TIMIT corpus is an English read speech corpus specifically designed for acoustic-phonetic studies \cite{garofolo1993darpa}. To measure the PER, we require phonemic transcriptions of the UASpeech utterances (reference). We used g2p-en\footnote{\url{https://github.com/Kyubyong/g2p}} for grapheme-to-phoneme conversion. The reference is compared to the VC utterances transcribed by the trained ASR.

\vspace{-0.2cm}
\subsection{Subjective evaluation protocols}

Subjective evaluation was carried out by naive listeners to assess the naturalness and similarity of samples (RQ1, RQ2). An additional evaluation was done by expert listeners to assess severity (RQ3). Contrary to the objective evaluations, we did not consider all conversion pairs (due to constraints in time and budget). Audio samples can be found online\footnote{\url{https://unilight.github.io/Publication-Demos/publications/n2d-vc}}.

\subsubsection{Severity}
We designed an AB evaluation study for evaluating severity (RQ3). In the study, 3 trained speech-language pathologists (SLPs) were asked to listen to two different synthesized utterances from two unknown speakers whom have different speech severity and select the synthesized speech sample that they perceived as being more pathological. We used four speaker pairs (see Table \ref{tab:severity}), two for each severity level. For each speaker pair, 20 utterances were rated. After rating the synthesized pathological speech samples, the experiment was repeated with the ground truth samples – as a  control for cases where we observe a reversal in the expected severity judgment in the VC speech samples.  So, in total, each SLP was asked to rate 80 utterances. A binomial test is performed to calculate significance.

\subsubsection{Naturalness}

In order to evaluate naturalness (RQ1), we followed the setup in  \cite{illa21_ssw} with a few modifications based on our previous findings. In our previous study, listeners rated the severity of the speech samples (instead of the naturalness) on a 5-point mean opinion score (MOS) scale. The results showed a flooring effect. Therefore, in this experiment, we increase the resolution of the MOS-scale to have increments of $0.5$. 
The questionnaire starts with an explanation of what we mean with naturalness, followed by an example of natural, normal and pathological (low severity) speech. The respondents were instructed to rate these both as 5 (highly natural). The stimuli consisted of 13 utterances for both pathological speakers of each severity (low, high, mid), leading to a total of 78 utterances.
Subsequently, the experiment was repeated with the ground truth samples. The utterances were rated by 30 native American English listeners. A Wilcoxon signed-rank test is performed to calculate significance.

\subsubsection{Similarity of the voice with the source normal speaker}
\label{section:method_subjective_similarity}
For the similarity (RQ2) evaluation, we follow the protocol in \cite{illa21_ssw}.
Listeners are presented a converted sample and a reference sample, and are asked to judge whether the two samples are uttered by the same speaker.
In short, the evaluation is AB similarity study where the source speaker is a pathological speaker, the target speaker is the control speaker.
The reference speech is either from the source (Similarity to source) or the target (Similarity to target).
We selected three pathological speakers (M04, M11, M10) which have deemed to have recognisable characteristics in our previous study \cite{illa21_ssw}. Furthermore, we randomly sampled (without replacement) two control speakers for each pathological speaker. The test were done by 5 naive Americna English listeners.
A binomial test is performed to calculate significance.

\begin{table}[t]
\caption{Objective evaluation results.}
\resizebox{\columnwidth}{!}{%
\begin{tabular}{l | r r r r r r | r r}
\toprule
& \multicolumn{2}{c}{\textbf{High}} & \multicolumn{2}{c}{\textbf{Mid}} & \multicolumn{2}{c|}{\textbf{Low}} & \\
\cmidrule(lr){2-3} \cmidrule(lr){4-5} \cmidrule(lr){6-7}
& M08 & M10 & M05 & M11 & M04 & M12 & $r$ & $r_{GT}$ \\
\midrule
P-STOI VTN &  0.73 &  0.75 &  0.62 &  0.60 &  0.58 &  0.45 & 0.88 & 0.89 \\
P-ESTOI VTN &  0.37 &  0.37 &  0.20 &  0.16 &  0.09 &  0.08 & 0.93 & 0.90 \\
P-STOI VTN-VAE &  0.73 &  0.75 &  0.62 &  0.63 &  0.61 &  0.45 & 0.84 & 0.89 \\
P-ESTOI VTN-VAE &  0.37 &  0.35 &  0.21 &  0.19 &  0.12 &  0.06 & 0.94 & 0.90 \\
\midrule
PER VTN &  58.7 &  55.1 & 84.1 & 71.8 &  79.6 &  103.4 & 0.83 & 0.70  \\
PER VTN-VAE &  62.9 &  59.3 &  106.3  &   76.2 &  81.2  &  120.0 & 0.68 & 0.70 \\ 
STER & 7.0 & 7.0  & 42.0  & 38.0 & 98.0 & 92.6 & 1.0 & -- \\ 
\bottomrule
\end{tabular}
\label{table:per}
}

\end{table}

\begin{table}[]
\centering
\caption{Mean opinion score results of the naturalness test with 95\% confidence intervals. Columns correspond to the intelligibility level, and rows correspond to ground truth (GT) and synthetic (VC) results. Higher is better.}
\label{tab:naturalness}
\begin{tabular}{lcccc}
\toprule
 & \multicolumn{1}{l}{\textbf{Normal}} & \multicolumn{1}{l}{\textbf{High}} & \multicolumn{1}{l}{\textbf{Mid}} & \multicolumn{1}{l}{\textbf{Low}} \\
\midrule
\textbf{GT} & 3.93 $\pm$ .54 & 3.92 $\pm$ .54 & 2.86 $\pm$ .89 & 2.32 $\pm$ 1.16 \\
\textbf{VC} & - & 2.70 $\pm$ .95 & 2.28 $\pm$ 1.03 & 1.94 $\pm$ 1.21 \\
\bottomrule
\end{tabular}%
\vspace{-0.3cm}
\end{table}


\begin{table}[t]
\caption{Results of the similarity AB experiments with 95\% confidence intervals. } 
\vspace{-0.5cm}
\begin{center}
\begin{tabular}{lll}
\toprule
{} &    Similarity to target & Similarity to source \\
\midrule
M04$\rightarrow$CM05 &  20\% $\pm$ 10\% &  32\% $\pm$ 12\% \\
M11$\rightarrow$CM09 &  37\% $\pm$ 13\% &  43\% $\pm$ 13\% \\
M10$\rightarrow$CF03 &  55\% $\pm$ 13\% &    8\% $\pm$ 7\% \\
M04$\rightarrow$CM04 &  33\% $\pm$ 12\% &  27\% $\pm$ 12\% \\
M11$\rightarrow$CM10 &  23\% $\pm$ 11\% &  32\% $\pm$ 12\% \\
M10$\rightarrow$CF02 &  48\% $\pm$ 13\% &   10\% $\pm$ 8\% \\
\midrule
Ideal &  100\% & 0\% \\
\bottomrule
\end{tabular}
\label{table:similarity}
\end{center}
\vspace{-0.5cm}
\end{table}

\section{Evaluation results}

\subsection{Objective evaluations}



\subsubsection{P-STOI/P-ESTOI}

The second block of Table~\ref{table:per} summarizes the results of the P-STOI/P-ESTOI analyses. In the VTN stage, the obtained correlation between the P-STOI/P-ESTOI and the STER are similar to the ones one would obtain with the GT ($r_{GT}$). Therefore, in the VTN stage the severity is well captured. In the VTN-VAE stage, the P-STOI correlation decreases from 0.88 to 0.84, while the P-ESTOI slightly increases from 0.93 to 0.94, which is a bit higher than ($r_{GT}$). This latter change can be explained as follows: the frame-based VAE model does not change the temporal aspects of the signal but rather the spectral aspects, for which the P-ESTOI has a higher sensitivity.




\subsubsection{Phoneme error rate}

The VTN PER results in Table \ref{table:per} show higher correlation with the STER than the GT,  which indicates that we can mimic the severity aspects of the pathological speech in the first stage. However, PER VTN-VAE results are decreased compared to the PER VTN. This is probably because the VAE stage causes a naturalness degradation.


\subsection{Subjective evaluations}

\begin{table}[t]
\centering
\caption{Percentage of ``correct'' answers in the AB severity tests for the ground truth samples and the different stages of the architecture. *** is $p < 0.001$; * $p < 0.05$}
\label{tab:severity}
\resizebox{\columnwidth}{!}{%
\begin{tabular}{lccccc}
\toprule
\textbf{Speaker pairs} & \multicolumn{1}{l}{\textbf{Ground truth}} & \textbf{VTN} & \multicolumn{1}{l}{\textbf{VTN-VAE}} & \multicolumn{2}{c}{\textbf{Severity}}   \\ 
\midrule
\textbf{M04 vs M05} & 95\% *** & 85\% *** & 53\% & Low & Mid \\
\textbf{M05 vs M08} & 90\% *** & 95\% *** & 80\% *** & Mid & High \\
\textbf{M12 vs M11} & 93\% *** & 85\% *** & 75\% * & Low & Mid \\
\textbf{M11 vs M10} & 98\% *** & 95\% *** & 68\% * & Mid & High \\
\bottomrule
\end{tabular}%
\label{table:severity}
}
\vspace{-0.5cm}
\end{table}

\subsubsection{Naturalness}

Table~\ref{tab:naturalness} shows the MOS results. First, similar to our previous study \cite{illa21_ssw}, we observe that with decreasing intelligibility, naive listeners perceive the heard speech increasingly unnatural -- even in the case of ground truth samples. Second, the ground truth samples are consistently rated as more natural then the  converted ones ($p < 0.001$).
Although, these results are not directly comparable to \cite{illa21_ssw}, we note that we've observed overall higher MOS values. We suggest that the use of seq2seq models contributed to this improvement, and such quality is sufficient for further investigation. 


\subsubsection{Similarity}

Table \ref{table:similarity} describes the identity preservation ability of the VC framework. We can see that the Similarity to source column has less than 50\% similarity for all speaker pairs, therefore we can conclude that the VC can successfully ignore the pathological source speaker's characteristics. However, we can also see from the Similarity to Target column that (except from the M10$\rightarrow$CF03) none of the VC samples have more than 50\% similarity to the target.
Such results emphasize the ``unobtainable ground truth ''difficulty faced by the model, as described in Section~\ref{sec:intro}. Meanwhile, this also points out that improving speaker similarity is an important future work, as this problem was also present in \cite{dvc-vtn-vae}.

\vspace{-0.2cm}
\subsubsection{Severity}

Table \ref{table:severity} lists the percentage of ``correct'' answers in the AB severity test done with the SLPs. 
On average, the SLPs always perceived the more severe speakers as more severe (each entry in Table \ref{table:severity} is over 50\%). In the first VTN stage, no more than 10\% decrease in "correct" answers is observed in the severity recognition compared to the ground truth. Furthermore, the ratio of "correct" severity decisions slightly increased in the case of the VTN M05 vs M08 pair. This indicates that the VTN simulates the severity aspects well. After the second VTN-VAE step, we see a decrease in ``correct'' answers, which means that in the case of speaker-specific samples, the SLPs made more errors in indicating which of the two samples had a worse severity, possibly because the severity difference is less perceivable for the SLPs due to the additional distortion caused by the VAE.

\vspace{-0.2cm}
\section{Conclusions}

In this paper, we proposed a novel two-stage framework for N2D VC. We evaluated the proposed method on UASpeech \cite{UASpeech}, and the method achieved good naturalness results, was able to mimic the severity characteristics in a linear way according to three speech language pathologists, while being able to convert away from the pathological source speaker's characteristic. In the future, we will focus on improving the preservation of the normal source speaker identity.

\noindent\textbf{Acknowledgements}
We would like to thank Lisette van der Molen, Klaske van Sluis, and Marise Neijman for participating in the severity experiment. All questionnaire participants were remunerated justly (7.50GBP/hour) in each experiment.
B.M.H. is funded through the EU’s H2020 research and innovation programme under MSC grant agreement No 766287. The Department of Head and Neck Oncology and Surgery of the Netherlands Cancer
Institute receives a research grant from Atos Medical (H\"orby, Sweden), which contributes to the existing infrastructure for quality of life research.
This work was partly supported by JSPS KAKENHI Grant Number 21J20920, JST CREST Grant Number JPMJCR19A3, and AMED under Grant Number JP21dk0310114, Japan.


\newpage
\bibliographystyle{IEEEbib}
\bibliography{ref}

\end{document}